\documentclass{article}
\usepackage{arxiv}

\usepackage[utf8]{inputenc} % allow utf-8 input
\usepackage[T1]{fontenc}    % use 8-bit T1 fonts
\usepackage{hyperref}       % hyperlinks
\usepackage{url}            % simple URL typesetting
\usepackage{booktabs}       % professional-quality tables
\usepackage{amsfonts}       % blackboard math symbols
\usepackage{nicefrac}       % compact symbols for 1/2, etc.
\usepackage{microtype}      % microtypography
\usepackage{lipsum}
\usepackage{graphicx}
\graphicspath{ {./images/} }
\usepackage{amsmath}
\usepackage[numbers]{natbib}
\usepackage{dirtytalk}
\usepackage{subfig}

\title{Learning a Perceptual Manifold with Deep Features for Animation Video Resequencing}

\author{
 Charles C.Morace \\
   \And
 Thi-Ngoc-Hanh Le \\
  \And
 Sheng-Yi Yao \\
 \And
 Shang-Wei Zhang \\
  \And
  Tong-Yee Lee\\
  Department of Computer Science and Information Engineering\\
  National Cheng-Kung University\\
  Taiwan\\
  \texttt{tonylee@mail.ncku.edu.tw} \\
}

\begin{document}
\maketitle
\begin{abstract}
We propose a novel deep learning framework for animation video resequencing. Our system produces new video sequences by minimizing a perceptual distance of images from an existing animation video clip. To measure perceptual distance, we utilize the activations of convolutional neural networks and learn a perceptual distance by training these features on a small network with data comprised of human perceptual judgments. We show that with this perceptual metric and graph-based manifold learning techniques, our framework can produce new smooth and visually appealing animation video results for a variety of animation video styles. In contrast to previous work on animation video resequencing, the proposed framework applies to wide range of image styles and does not require hand-crafted feature extraction, background subtraction, or feature correspondence. In addition, we also show that our framework has applications to appealing arrange unordered collections of images.

\keywords{deep learning \and deep features \and manifold learning \and animation video resequencing
}
\end{abstract}

\section{Introduction}
From its beginnings, animation has brought to life the creative potential of the human mind. It has developed into a dominant visual storytelling tool, and today there exists a wealth of archived animation video sequences created with both traditional and modern computer animation techniques. The visual style of animation video sequences is diverse, from stop-motion and three-dimensional photo-realistic renderings to cartoon illustrations and line-sketches. Although many techniques have been developed to ease the computer animation pipeline, production is still an arduous process, in large part to the complexity of digital characters, environments, and motion. The goal of this paper is to regenerate new video resequence for general and diverse animation video source. Animators can efficiently and interactively regenerate new animations according to their desire. Therefore, it reduces the complexity and timing/expense cost in creating animations. \par 

\begin{figure}[hbt!]
  \centering
  \includegraphics[width=0.65\textwidth]{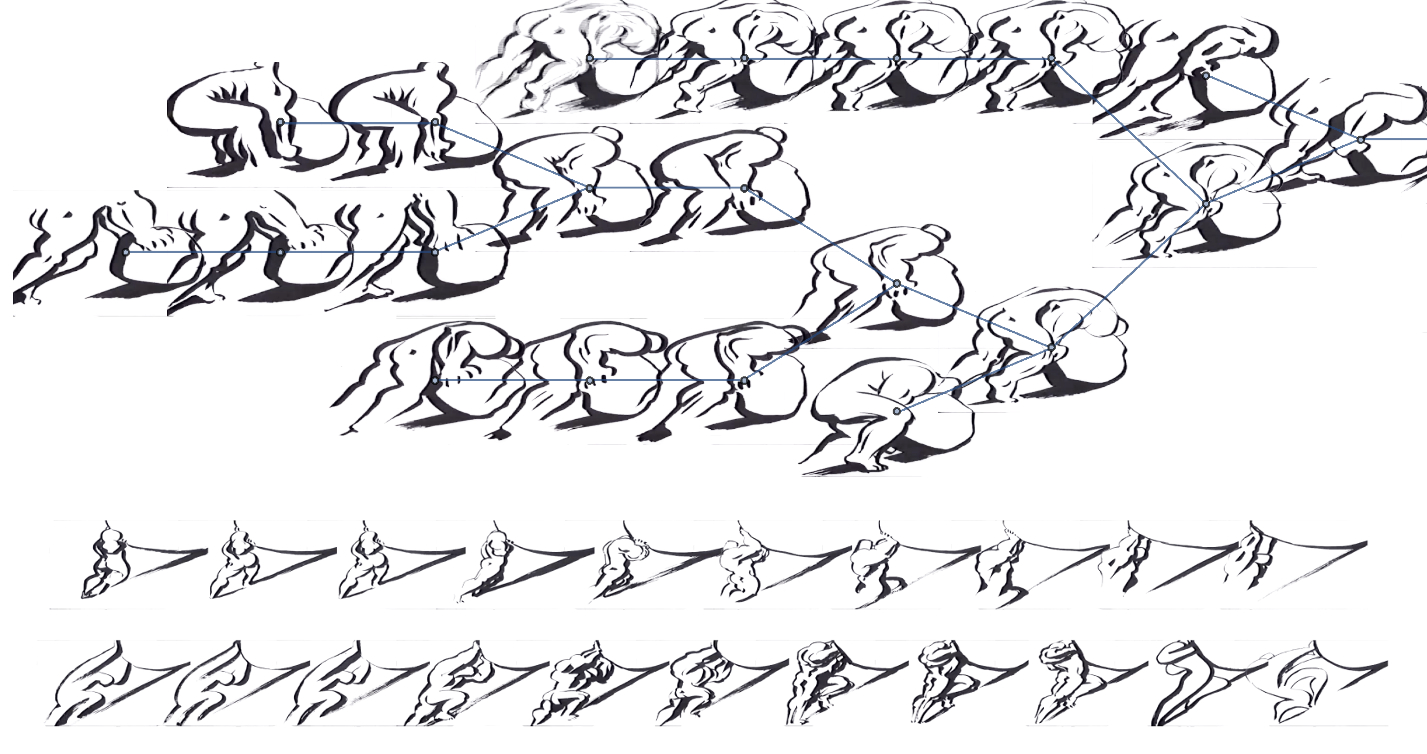}
  \caption{\small An example of the manifold topology and two animation video resequences generated from a collection of unordered animation images with the proposed method.}
  \label{fig_1}
\end{figure}

Previous works on animation resequencing \cite{de2004cartoon, yu2012combining, yu2012semisupervised} have focused on hand-craft feature extraction techniques to measure texture and shape similarity for a cartoon style animation and require background subtraction, segmentation, and other image processing techniques. Motivated by this, we propose a framework for animation video resequencing which can be applied to a broader range of image styles and does not require hand-craft feature extraction, background subtraction, or feature correspondence. Given the abundance of animation data which is currently available, our proposed resequencing framework generates new animations from an existing animation video clips. \par

The proposed framework learns a topological manifold of images where paths on the manifold represent smooth and visually plausible animation video frame sequences. To demonstrate our proposed framework to be general enough to handle a variety of animation video styles as source data, including photorealistic, non-photorealistic, styzied, and line-sketches we experimentally evaluate our method using different kinds of image and animation data. Also, given selected key-frames, our animation video resequencing method can create smooth and appealing user-controlled animation video.\par

In this paper, we utilize the activations of deep convolutional neural networks for smooth image sequencing and animation video resequencing. Given a pre-trained CNN and selected activation layers, we learn a perceptual similarity metric which reflects the perceptual judgments of humans. Then, from a collection of images and their pair-wise perceptual distances, we generate smooth new animation video sequences by traversing paths and cycles in an estimated perceptual manifold. To the best of our knowledge, we are the first to apply deep features to the problem of animation video resequencing. Our technical contributions can be summarized in the following issues:
\begin{itemize}
    \item [$\bullet$] We combine deep feature extraction and a perceptual similarity metric with a graph-based manifold learning technique to generate new and smooth animation video sequences.
    \item [$\bullet$] We implement our method with two well-known deep learning architectures, VGG \cite{simonyan2014very}, and AlexNet \cite{krizhevsky2017imagenet}, and perform an experimental evaluation of the deep features learned by these architectures.
    \item [$\bullet$] We give a quantitative comparison of our animation video resequencing results with other image similarity metrics including $L_2$ distance in image space, $L_2$ distance of the bottleneck layer activations of a denoising autoencoder \cite{vincent2010stacked}, and results obtained by traditional manifold learning techniques, locally Linear Embedding \cite{roweis2000nonlinear} and Isomap \cite{tenenbaum2000global}.
    \item[$\bullet$] We demonstrate that our method can facilitate several applications such as creating image layouts and video synthesis.
    \item[$\bullet$] To the best of our knowledge, this is the first work that shows our framework can generally handle different kinds of animation styles without extra efforts on different data.
\end{itemize} 

We organize the remainder of this paper as follows. In section 2, most previous research in this sequencing domain are reviewed. In section 3, we introduce the overview of our proposed system. In section 4, the methods used in our system are described. In section 5, our experimental results and evaluations are presented and our additional applications follow. The conclusions and our future work are presented in the last section.

% ================================================= Related work ===========================================

\section{Related work}
Smooth image sequencing is the key to producing a visually pleasing animation video. Most previous research divides this sequencing problem into two distinct steps. The first step is to establish a suitable distance measure for the similarity between the input images. The second step is to determine an optimal sequence according to the similarity measure defined in the first step. While the earliest work \cite{schodl2000video} used simple $L_2$ distance in the original image space as a similarity measure, more recent works have focused on feature extraction and dimension reduction techniques to measure higher-level features of shape \cite{ling2007shape}, appearance \cite{fried2017patch2vec}, pose \cite{osadchy2007synergistic}, and motion \cite{holden2015learning}.\par

In the proposed method, we use the activations of deep convolutional neural networks for feature extraction and a metric inspired by the Learned Perceptual Image Patch Similarity (LPIPS) metric proposed by \cite{zhang2018unreasonable} to measure the perceptual distance of images. Although PSNR and SSIM \cite{hore2010image} are mentioned as a well-known perceptual metrics, they are widely used to measure the similarity between two images. The goal of our work is to learn a \say{perceptual manifold} with deep features, LPIPS metric is suitable for perceptual similarity across deep visual representations \cite{zhang2018unreasonable}.\par

\subsection{Feature Extraction and Non-linear Dimension Reduction}
Patch2Vec \cite{fried2017patch2vec} proposed a novel learning framework for image patch embedding, where an embedding is learned so that $L_2$ distance in the embedding space provides a useful measure for high-level features of texture dissimilarity. They trained a convolutional neural network (CNN), using a segmentation dataset and triplet loss function to map image patches having the same texture to points which are nearby in the embedding space while mapping image patches having other textures as far away as possible. The proposed method takes a similar approach by using deep convolutional networks to extract an optimal set of image features. However, instead of learning a perceptual metric with a triplet loss, we use the LPIPS metric trained on perceptual judgments of humans.\par

\citet{osadchy2007synergistic} proposed a view-independent Energy-Based Model that simultaneously detects faces and estimates head pose. They train a CNN to map images containing faces to points on a lower-dimensional face-manifold. After training, if the CNN maps an image to a point close to the manifold then its pitch, roll, and yaw are estimated by the position of the point projection onto the manifold. However, for general animation sequencing, there does not have a related energy-based model since the motion of a variety of characters and scenes must be estimated.\par

\citet{holden2015learning} used a convolutional autoencoder for learning a manifold of human motion. They trained their proposed CNN with a motion capture data consisting of time-series of human-poses, and each convolution layer of the CNN performs one-dimensional convolution over the temporal domain. Since their CNN is trained using motion capture data, it is not suitable for feature extraction of images. \par

\citet{yu2012combining} described methods for cartoon retrieval and clip synthesis using a multi-feature distance function and a partially user-labeled database of cartoon characters to construct a lower dimensional feature space with a sparse transfer learning technique. Cartoon Textures \cite{de2004cartoon} proposed a feature-distance based on the shape \cite{huttenlocher1993comparing}, appearance, and the temporal ordering of an input cartoon sequence, and then utilize the manifold learning technique Spatio-Temporal Isomap (ST-Isomap) to recover a lower-dimensional embedding of the input. Unlike the proposed method, ST-Isomap requires an initial ordering for the input images and thus does not apply to unordered collections of images. Moreover, ST-Isomap, and traditional manifold learning techniques like Locally Linear Embedding (LLE) \cite{roweis2000nonlinear} and Laplacian Eigen Maps (LEM) \cite{belkin2003laplacian} required predetermined parameters including the dimension of the manifold and the number of neighbors of each input image. These parameters require fine-tuning for each collection of input images. Cartoon Textures \cite{de2004cartoon} and the methods proposed by \cite{yu2012combining, yu2012interactive} all use hand-crafted feature-extraction methods specific to cartoon images and require much preprocessing including segmentation of a character and pairwise computation of the Hausdorff distance. Furthermore, the Cartoon Textures \cite{de2004cartoon} method relied on knowledge of the ordering of the original input sequence and the methods of \cite{yu2012combining, yu2012interactive} required user labeled data to construct the embedding to measure image similarity. In contrast, the method proposed in this paper, requires no user-labeling, no segmentation, and works with a variety of image styles.

\subsection{Sequential Ordering of Images}
Determining a sequential ordering of images is usually posed as a path-finding problem in a weighted graph, where nodes correspond to images and transition costs are based on the image dissimilarity measure and possibly other criteria such as path smoothness, temporal ordering of the input sequence, or user-control. Previous work considers transition costs in complete graphs and nearest neighbor graphs.\par

Video textures \cite{schodl2000video} is a video-based rendering technique. They apply Q-learning \cite{kaelbling1996reinforcement} to generate a video sequence of arbitrary length with similar dynamics to the input video. The method produces convincing results when the input video has repetitive motion or unstructured stochastic motion but will fail for complex structured motion like full body human motion. This limitation is a consequence of using $L_2$ distance on the raw pixels of the images which cannot sufficiently measure similarity in high-level features of motion. To ameliorate this issue, Shödl and Essa \cite{schodl2001machine, schodl2002controlled} trained a linear binary-classifier from manually labeled training data based on six hand-crafted features. Images deemed unacceptable by the linear classifier are not considered for transitions, and for the other images, the similarity measure is the linear classifying function. Shödl and Essa \cite{schodl2001machine} applied a beam-search technique to obtain the optimal sequence. To improve the sequencing results of their beam search, Shödl and Essa \cite{schodl2002controlled} considered the temporal information of the original input sequence in the transition cost and adopt a greedy hill-climbing optimization, which starts with a random image sequence and iteratively changes subsequences which lower the total path cost. Contrast to these previous methods, our method applies to unordered and unlabeled collections of images. \par

Unlike the previously discussed techniques which sequence photo-realistic images, Cartoon textures \cite{de2004cartoon} and the works by \cite{yu2012combining, yu2012interactive} synthesize cartoon animations. \citet{yu2012interactive} used a greedy method, choosing a random cartoon image as the first frame, and then choosing the most similar image, measured in the low dimensional subspace, for each subsequent frame. Cartoon Textures and \citet{yu2012combining} synthesized new animations by finding the shortest paths in a graph constructed by the ST-Isomap and Isomap manifold learning algorithms, respectively. It is important to note that traditional manifold learning algorithms such as ST-Isomap and Isomap do not construct the graph automatically and require defined neighbor relations for the input data and the dimension of the embeddings beforehand, which is difficult to estimate. In our framework, we do not compute an explicit embedding and can adapt to different CNN architectures. Also, we automatically determined neighbor relations by minimizing the perceptual distance of the input data. Therefore, our method does not require fine-tuning for each input collection like traditional manifold learning techniques.

% ================================================= System overview ===========================================

\section{System overview}
\begin{figure}[hbt!]
  \centering
  \includegraphics[width=\textwidth]{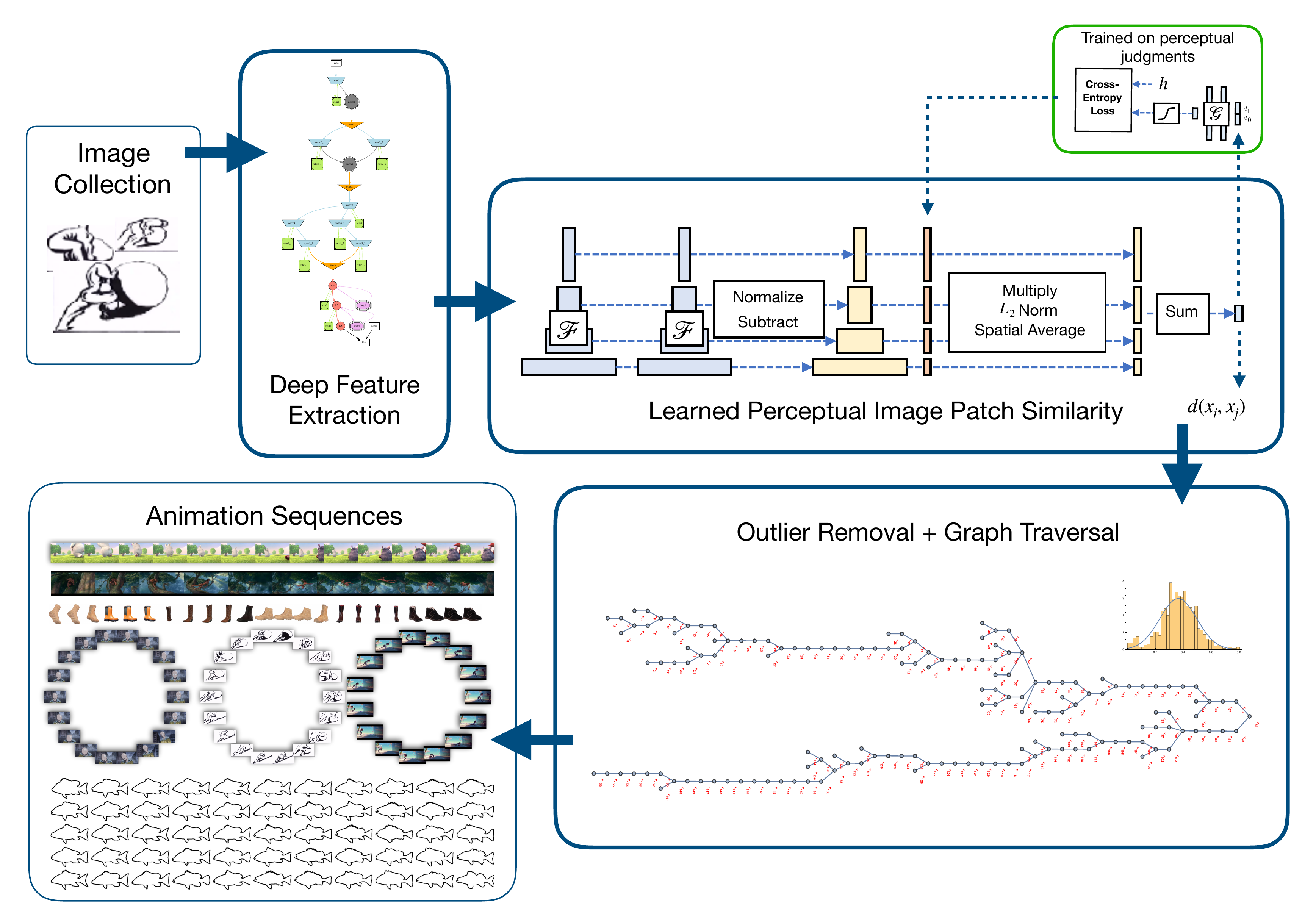}
  \caption{\small System overview of the proposed method.}
  \label{fig_2}
\end{figure}

We outline the system overview of our proposed framework in Figure \ref{fig_2}. The input to our system is a collection of images $\mathbf{X}$ and a trained CNN $\mathbf{F}$. The CNN serves as an image feature extractor, and we learn a \say{perceptual distance} by training another neural network $\mathbf{G}$ on a dataset comprised of perceptual judgments of humans. In our implementation, we test the activations of AlexNet and VGG trained for image classification. However, features extracted by other CNNs trained for tasks other than classification are also useful for measuring perceptual distance \cite{zhang2018unreasonable} and could be incorporated into our system. So, after extracting the deep features from each image, we compute the pairwise perceptual distance of each image in the input collection using the LPIPS metric proposed by \cite{zhang2018unreasonable} Once the perceptual distance is learned, the proposed system can create:
\begin{itemize}
    \item [$\bullet$] a path sequence which uses all images in the collection given start and terminal frames;
    \item[$\bullet$] a cycle animation which uses all or some subset of images in the collection;
    \item[$\bullet$] or a path animation with smooth in-between images given a set of key-frames.
\end{itemize}

For some collections of images, it may not be possible to obtain a smooth animation sequence using every image in the collection. For input collections obtained from densely sampled videos, we can assume there exists at least one smooth sequence which uses all images in the input collection, i.e., the original animation. However, if the images come from sparsely sampling an existing animation video or from an unordered collection of images, it may not be possible to resequence all of the images into a smooth sequence. Therefore, we also detect and prune outliner images from the input collection by fitting the perceptual distance of nearest neighbors to an optimal probability distribution using maximum likelihood estimation.

Then, from the pairwise perceptual distances, we construct a complete graph where each image corresponds to a node, and the weight of each edge is equal to the perceptual distance of adjacent images. To generate an optimal animation sequence through all of the input images, we compute the shortest Hamiltonian path from starting, and terminal frames assigned by the user or generate a looping animation by computing shortest tours. In key-frame path finding, we compute a minimal spanning tree (MST) of the complete graph and generate animation sequences by traversing paths in the MST. Figure \ref{fig_2} shows an overview of the system.

Section 4.1 describes the feature extraction details, section 4.2 describes the procedure for computing perceptual distance, section 4.3 describes the method for automatic outliner detection and removal, and section 4.4 describes animation resequencing.

% ================================================= Method ===========================================

\section{Method}
\subsection{Deep feature extraction}
To compute the perceptual distance, we first extract features with a trained convolutional neural network (CNN) $\mathbf{F}$. Image features for an image $x_i \in X$ are a set of activations, $\{\hat{y}^l_i \in \mathbb{R}^{w_lh_lc_l}\}^L_{l=1}$, obtained from the activations of selected layers after applying the image $x_i$ to the CNN $\mathbf{F}$, where $L$ is the total number of selected activation layers, and $w_l$, $h_l$, and $c_l$ are the dimensions of a selected activation layer $l$'s output. Thus, the size of features and the speed at which that are extracted depends on the architecture of the network architecture of $\mathbf{F}$.

In our implementation, we test two off-the-shelf networks, namely, VGG \citep{simonyan2014very} and AlexNet \cite{krizhevsky2017imagenet}. Both two networks are trained on ImageNet dataset \cite{russakovsky2015imagenet} and have excellent performance in image classification, object detection, etc. Besides, there are number of deep learning-based studies in this problem domain adopting them to extract features in their system. Thus, these two pre-trained networks are suitable to adequately extract features from images. In each backbone, we remove the later layers, fully connected, and utilize the first five activation layers as feature extractors. While VGG and AlexNet are trained for classification of natural images, our experimental results show that their activations are also useful for re-sequencing non-photorealistic image styles used in animation.

\subsection{Perceptual Distance}
In this section, we briefly summarize the perceptual distance used for generating animation sequences. We use the LPIPS metric proposed by \cite{zhang2018unreasonable}. Given a trained convolutional network $\mathbf{F}$ and two images $x_i$ and $x_j$, we extract the activations from $L$ selected layers and unit-normalize in the channel dimension to obtain features $\hat{y}^l_i$ and $\hat{y}^l_j$ for each selected layer $l \in \{1, \dots, L\}$. To compute the perceptual distance $d(x_i,x_j)$, we scale the difference of activations element-wise by learned \say{perceptual callibration} weight tensors $w_l$, compute the $L_2$ norm, average spatially, and sum over all layers. This distance is expressed as follows:

\begin{equation}
    d(x_i, x_j) = \sum_l \frac{1}{H_l W_l} \sum_{h,w} \parallel w_l \odot (\hat{y}^l_{ihw} - \hat{y}^l_{jhw} \parallel^2,
\end{equation} where the weights $w_l$ are learned by a small network $\mathbf{G}$ trained to predict perceptual judgement $h$ from distance pairs $(d_0, d_1)$ where $d_0 = (x, x_0)$, $d_1 = (x, x_1)$, $x$ is a reference images, and $x_0$ and $x_1$ are distorted images of $x$. The judgement $h \in (0, 1)$ is determined based on the proportion of humans that perceived the image $x$ to be more similar to $x_1$ than $x_0$ and weights $w_l$ are obtained by minimizing a cross entropy loss function which is formulated as:

\begin{equation}
\begin{split}
    L(x, x_0, x_1, h) &= -h\log G\big(d(x, x_0), d(x, x_1)\big) \\
    & - (1-h)\log\big(1-G\big(d(x, x_0), d(x, x_1)\big)\big)
\end{split}
\end{equation}

The perceptual judgments are obtained from the publically available Two Alternative Forced Choice (2AFC) dataset collected by \cite{zhang2018unreasonable}. For additional details, we refer the reader to the original paper \cite{zhang2018unreasonable}. For each pair of images $x_i$ and $x_j$ in the input collection, we use the LPIPS metric in equation 1 to compute a \say{perceptual distance}.\par

In section 5.2 we show a comparative analysis of animation results obtained with the LPIPS metric and other image similarity metrics, including $L_2$ in image space, $L_2$ in a denoising autoencoder’s bottleneck activation feature space, two traditional manifold learning methods’ embeddings, and the cosine distance of the same deep features used with LPIPS.\par

\subsection{Outliner Detection and Removal}
Images which have a large perceptual distance from all other images in the input collection may negatively affect the smoothness of the resequenced animation result. Thus to maintain the smoothness of the generated sequence, we remove outliner images which have a large perceptual distance from their nearest neighbors. A naive approach would be to simply threshold the perceptual distance of nearest neighbors in the complete graph. However, a constant threshold value cannot adapt to disparate input data. Therefore, we fit the perceptual distance of nearest neighbors to a probability distribution to detect and remove outliner images. \par

Let $X_i$ be a random variable equal to the average perceptual distance of image $x_i$ and its nearest neighbors $x_{i(j)}$ for $j \in {1, \dots, K}$.
\begin{equation}
    X_i = \frac{1}{K} \sum^K_{j-1} d(x_i, x_{i(j)}),
\end{equation} In our implement, we choose the number of nearest neighbors to be $K=5$. \par

To find the most likely distribution, we estimate the parameters of the generalized gamma probability distribution function \cite{stacy1962generalization} given the samples $X_i$ for $i \in {1, \dots, N}$ where $N$ is the total number of images in the input collection. The generalized gamma probability density is defined as,
\begin{equation}
f(x;\alpha, \beta, \gamma, \mu) \propto =\begin{cases}
			(x-\mu)^{\alpha \gamma-1} exp\big(-\big(\frac{x-\mu}{\beta}\big)^\gamma\big), & \text{if } x>\mu\\
            0, & \text{if } x<\mu,
		 \end{cases}
\end{equation}The parameters $\alpha$, $\beta$, $\gamma$, and $\mu$ are obtained with maximum likelihood estimation, i.e., by maximizing the log likelihood function of $f$ given the random samples $X_i$. Once the parameters are found, we calculate the 0.9 quantile value $T$ as a threshold and remove the $i-th$ column and $i-th$ row from the original distance matrix for any $X_i>T$ and update the complete graph.\par

We choose the generalized gamma function as a distribution because of its flexibility. We tested the Normal and Beta distributions but found that the generalized gamma distribution produced better fits to the sample histograms than other distribution models.\par

\subsection{Animation Resequencing}
From the perceptual distance matrix, we construct a complete graph $G(V,E$) where a node $v_i \in V$ corresponds to image $x_i$ and the weight of an edge $e_{ij}$ is $d(x_i,x_j)$. Initially, we view the complete graph as a crude approximation of a perceptual manifold. Traversing the complete graph would allow for large \say{perceptual jumps} since each pair of images are adjacent and a large perceptual distance of adjacent frames would result in an unsmooth animation sequence. To improve our estimation of the perception manifold, we find subgraphs which prune large edges from the complete graph and generate animations by traversing a modified graph structure. We consider three different types of subgraphs, the shortest Hamiltonian path \cite{cormen2009introduction}, the shortest Hamiltonian cycle \cite{cormen2009introduction}, and the minimum spanning tree (MST) \cite{cormen2009introduction}. \par

In this section, we formalize these problems and give additional details and justification for these methods. In general, verifying if a sequence is a shortest Hamiltonian path or shortest Hamiltonian cycle is an NP-complete problem \cite{garey1979computers}. For a given input collection with $m$-frames, verification requires an exhaustive search of all $m!$ permutations of the set $\{1,\dots, m\}$. For larger image collections, finding an exact solution quickly becomes infeasible. However, for a complete graph, the existence of a Hamiltonian path and Hamiltonian cycle is guaranteed, and many polynomial approximation algorithms with bounded error have been proposed \cite{laporte1992traveling}. In our implementation we use commercial software Mathematica \cite{Mathematica} to solve the shortest Hamiltonian path and Hamiltonian cycle problems. The MST, on the other hand, can be computed very efficiently using a greedy method such as Kruskal’s algorithm \cite{kruskal1956shortest}.

\subsubsection{Shortest Hamiltonian Path Sequence}
 For an input collection of images with $m$ frames, the shortest Hamiltonian path in the complete graph is the permutation of images $\psi$ in the set of images $\Psi$ which minimizes the total perceptual distance between adjacent frames:
 \begin{equation}
     min_{\psi \in \Psi} \sum^{m-1}_{i=1} d\big(x_{\psi(i)}, x_{\psi(i+1)}\big),
 \end{equation}
Optionally, a user can add constraints to the set of permutations so that the first frame in the Hamiltonian path has index $\psi(1) = s$ and the terminal frame has index $\psi(m) = t \neq s$. For input animations which do not contain cyclic motion, this method can be used to reconstruct the original animation sequence given $s=1$ and $t=m$.

\subsubsection{Shortest Hamiltonian Cycle Sequence}
\begin{figure}[hbt!]
  \centering
  \includegraphics[width=\textwidth]{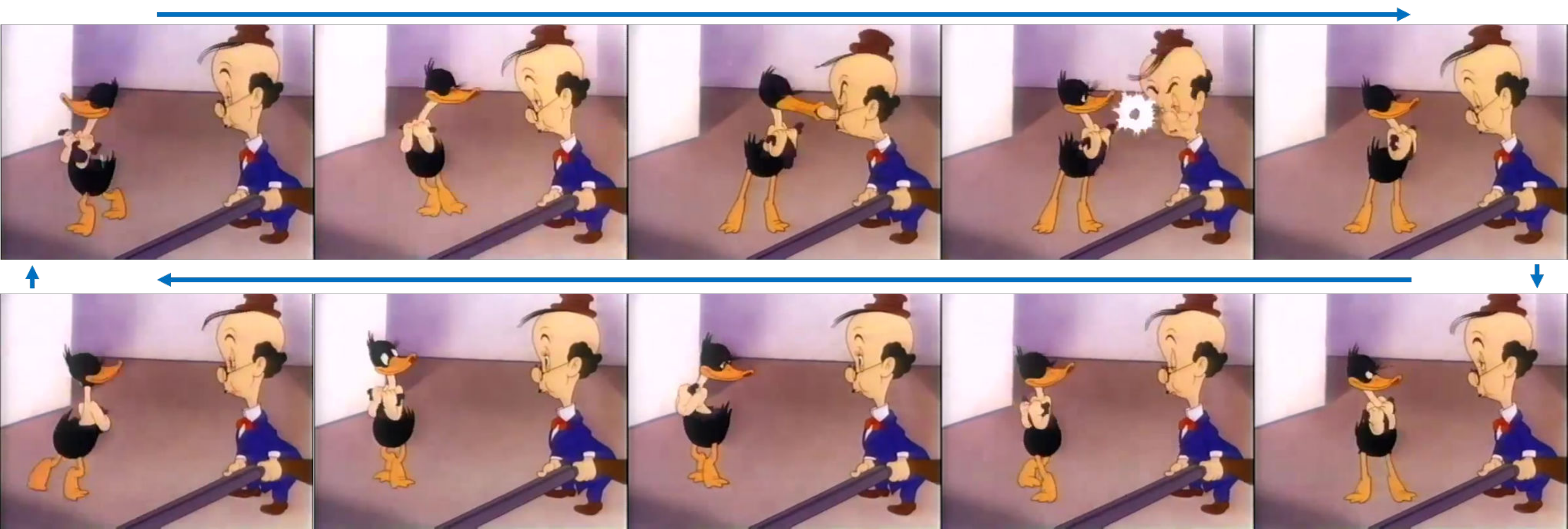}
  \caption{\small Results from uniformly sampling the shortest Hamiltonian cycle sequence. In this example, the sequence is generated from six images. The blue arrows navigate the starting frame (top-left frame)to the ending frame (bottom-left frame) in a cycle.}
  \label{fig_4}
\end{figure}

To compute a cyclic animation sequences, we compute the shortest Hamiltonian cycle of the complete graph. Finding the shortest Hamiltonian cycle is equivalent to the well-known traveling salesman problem, and corresponds to a cyclic permutation of images $\psi$ which minimizes the total perceptual distance between adjacent frames in set $\Psi$:
\begin{equation}
    min_{\psi \in \Psi} d\big(x_{\psi(1)}, x_{\psi(m)}\big) + \sum^{m-1}_{i=1}d\big(x_{\psi(i)},x_{\psi(i+1)}\big),
\end{equation}
The Hamiltonian cycle can generate looping sequences with continuously smooth motion which can be chained together to create a looping animations of arbitrary length. We show the result from uniformly sampling the shortest Hamiltonian Cycle Sequence in Figure \ref{fig_4}.

\subsubsection{Key-frame Path-finding}
In key-frame path-finding, we use paths in the minimum spanning tree (MST) to return temporally-coherent in-between frames given a set of key-frames by the user. Animators typically choose key-frames as the beginning and end points of a temporally coherent transition. Thus, for in-between sequencing, we would like a high level of confidence that in-between images remain close to the perceptual manifold and we hope to return a sequence of many temporally coherent images to the user.\par

Since all of the perceptual distances are positive, the MST is the minimum-distance subgraph which connects all of the images. Therefore our proposed method produces in-betweens by traversing the path from one key-frame node to another along an MST. The paths connecting key-frame nodes in an MST are well suited for finding in-between images since the distance between nodes is relatively small which gives us a higher level of confidence that the in-between images are temporally coherent. The MST also has the advantage of having the minimal set of edges for a path-connected graph containing each image in the input collection, thus reducing both the time and space complexity of path-finding.\par

With this method, users may create animations from any number of key-frames by computing paths between consecutive key-frames and combining the results. The user can also view a 2D linear embedding of the MST to see an overview of the entire dataset and help drive their decisions in key-frame selection. Figure \ref{fig_1} shows how a user can use the MST’s 2D linear embedding to choose key-frames and view the sequence of in-between images.

\section{Experimental Results}
We separate our experimental results and evaluation into five subsections. In section 5.1, we show the training details in our proposed framework. In section 5.2, we show some representative results of our framework and a variety of animation sequences generated with the Hamiltonian path, Hamiltonian cycle, and key-frame pathfinding methods. In section 5.3, we compare the effectiveness of applying the deep features and the LPIPS metric to animation resequencing with other image similarity metrics. We give a quantitative comparison between $L_2$ distance in image space, $L_2$ distance of the bottleneck layer of a denoising autoencoder, traditional manifold learning algorithms LLE and Isomap, cosine distance and the LPIPS metric \cite{zhang2018unreasonable} applied to the deep features extracted by VGG and AlexNet. In section 5.4, we show results of our framework applied to unordered image sets for image layout applications and in section 5.5 we discuss the main limitations of our framework.\par

In addition to the results presented here, please see our supplementary material and video for additional results and comparisons which are available on our project website:\newline \url{http://graphics.csie.ncku.edu.tw/ManifoldAnimationSequence}. 

\subsection{Training details}
\begin{figure}[hbt!]
  \centering
  \includegraphics[width=0.85\textwidth]{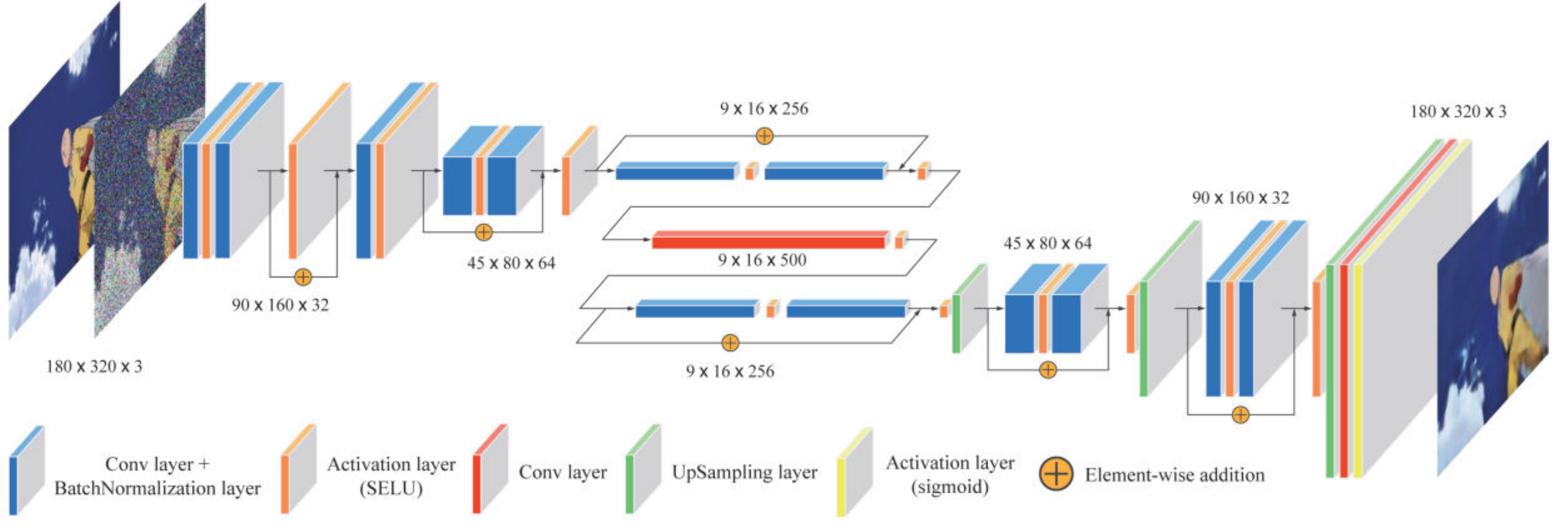}
  \caption{\small The architecture of the denoising autoencoder used in testing animation reconstruction.}  \label{fig_autoencoder}
\end{figure}

In our network architecture, we use both general convolutional layers and the concept of residual blocks \cite{he2016deep}. A residual block is a block of layers where the input to the block is added element-wise to the output of  the  block.  This technique helps  prevent  vanishing  gradients  which  is a common problem in training deep neural  networks. We use scaled exponential linear units (SELUs) \cite{klambauer2017self} as our activation functions except in the last layer where we apply a  sigmoid  function  to  guarantee  that  the  output  image pixel values are between zero and one. We  also use batch normalization layers \cite{ioffe2015batch} in our model to keep the values of tensors propagating in the network to have zero mean and unit variance. The benefits of the SELU activation functions and batch normalization layers are to train a deeper network and make training converge faster. Figure \ref{fig_autoencoder} gives additional details about the network architecture.\par

For training, we collect 20 Japanese cartoon animations, where each video is about 25 minutes long, and linearly scaled each frame to $w = 320$ pixels and a height $h = 180$ pixels to reduce training time. To avoid images which are nearly identical, we obtained the training images by uniformly sampling one out of every ten frames. In total, our training set and validation set consists of 60000 and 10000 images, respectively. We use $L_2$ distance to measures the error between the original images $X$ and the reconstructed image $\phi(X)$:
\begin{equation}
    L(X, \phi(X)) = \sum^h_{i=1} \sum^w_{j=1} \parallel X_{i, j} - \phi(X)_{i, j} \parallel^2,
\end{equation} where $X_{i, j}$ and $\phi(X)_{i, j} \in [0, 1]^3$ are the red, green, and blue components of the pixel($i,j$) in the original image and the reconstructed image, respectively. Then the optimal parameters of the encoder and decoder, are those which minimize a mean-square error loss across all iterations of the training process, where the batch size is set to 16. The initial parameters of the autoencoder, $\{\theta_0, \theta'_0\}$, are set by drawing samples from a truncated normal distribution similar to the technique described by Klambauer \cite{klambauer2017self}. Finally, we use the stochastic gradient descent algorithm ADAM \cite{kingma2014adam} to obtain the optimal solution.

\subsection{Animation Resequencing Results}
In this section, we show some representative results generated with our framework. To generate new animation sequences we collected test data by sampling frames from animation videos, extracting deep features from the first five activation layers of VGG, applying the LPIPS metric to each pair of images in the input collection, and resequencing the animations with the proposed outliner and graph traversal methods. \par

\begin{figure}[hbt!]
  \centering
  \includegraphics[width=\textwidth]{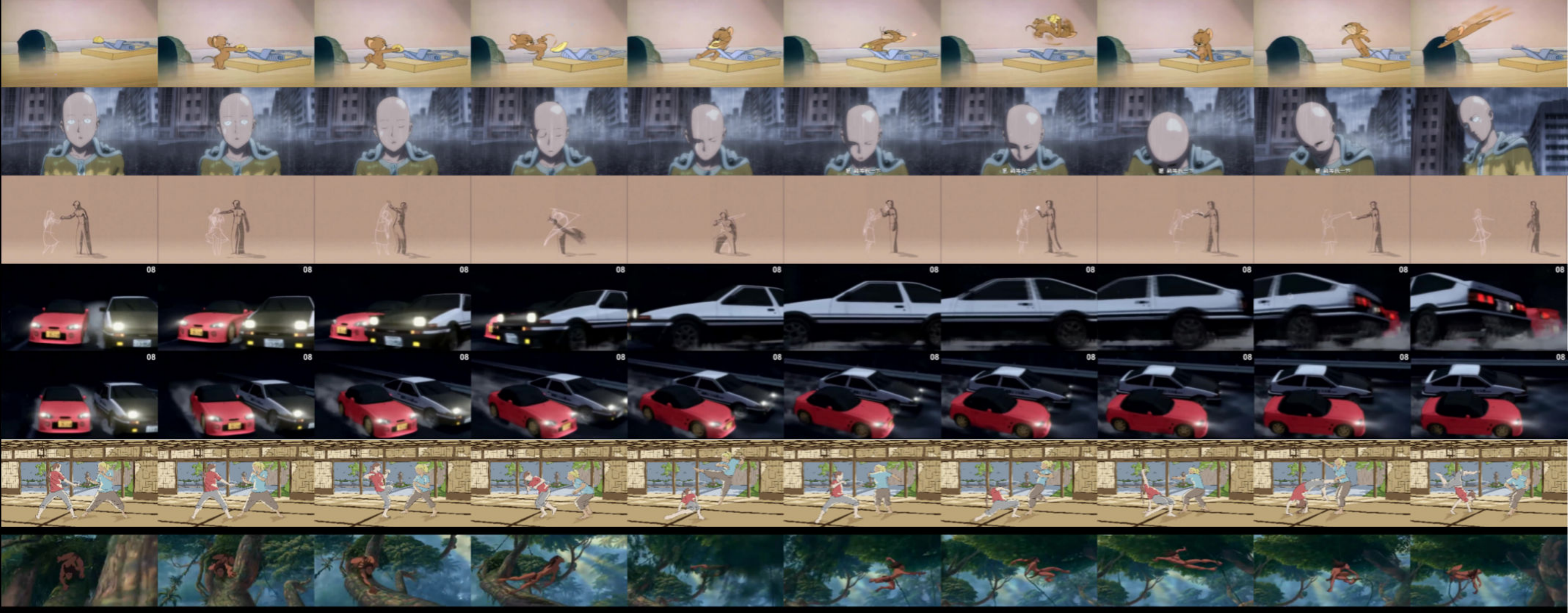}
  \caption{\small Results from uniformly sampling the shortest Hamiltonian path sequence.}
  \label{fig_3}
\end{figure}

Figure \ref{fig_3} shows uniformly sampled frames of sequences generated by computing the Hamiltonian path. We show the full sequences in our supplementary video as well as a comparison with animation results using features extracted from AlexNet and other feature extraction methods described in section 5.2. The Hamiltonian path often reconstructs the original animation if the user gives the initial and terminal frames as constraints. However, by using the outliner removal and other frame constraints it is possible to create new motion sequences which do not resemble the original input. \par

Our results show that Hamiltonian cycles can generate novel looping motion. A Hamiltonian cycle can create a pleasing looping effect, even when the input images come from an animation that is not originally a loop. Figure \ref{fig_4} shows a uniform sampling of the Hamiltonian cycle results. See our supplementary material and video for additional results.
Although we obtained many pleasing looking results, there are input data where the Hamiltonian cycle cannot immediately produce a smooth looping sequence. The proposed outliner removal method can improve outcomes in some of these cases. Another option is manually removing outliner images. One advantage of our system is that image layouts of the Hamiltonian cycle make it much easier to visually detect outliners and smooth subsequences.\par

\begin{figure}[hbt!]
  \centering
  \includegraphics[width=\textwidth]{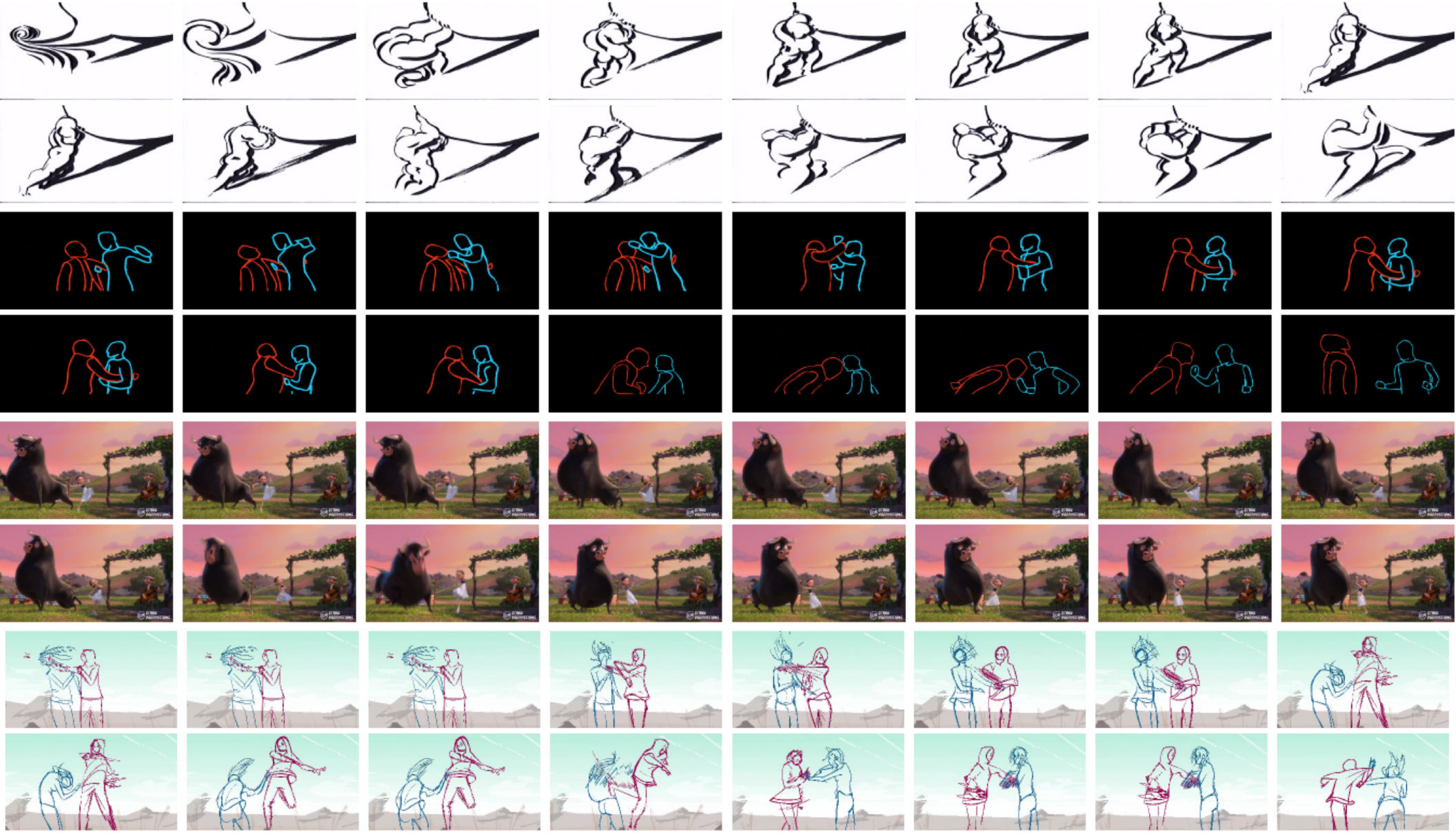}
  \caption{\small Results of our proposed key-frame method. The first and last frames are selected by a user and the in-between frames are generated by traversing the minimum spanning tree.}
  \label{fig_5}
\end{figure}

Figure \ref{fig_5} shows key-frame results generated with the proposed method. To create these results, we examined the MSTs to guide key-frame selection and return precisely six in-between images. In general, the user cannot directly control the number of in-between images returned for arbitrary key-frame selection. However, using the linear embeddings of the MST for visualization provides a useful way to select key-frames that produce the desired number of in-betweens. Figure \ref{fig_1} shows a portion of an MST’s 2D linear embedding and our supplementary material shows the full versions for all results shown in the paper.\par

The in-between frames generated by the proposed method are typically temporally-coherent for key-frames which have relatively short path distance in the MST, but as the path distance between key-frame nodes increase, so does the probability of unreasonable in-betweens. In practice, we do not consider this a significant draw-back since choosing additional intermediate key-frames can  avoid this issue.

\subsection{LPIPS Evaluation}
To evaluate the LPIPS performance for animation sequencing, we perform an animation reconstruction experiment with six related image similarity metrics. We collected 39 animations between 24 and 230 frames in lengths, shuffle the images, and attempt to reconstruct the original sequence by finding an optimal sequence which minimizes image dissimilarities between adjacent frames. The animations vary in style and content. We show original animations and the reconstructions for each image dissimilarity used for comparison in the supplementary video. \par

We compare the LPIPS metric of deep features of VGG and AlexNet with:
\begin{itemize}
    \item[$\bullet$] $L_2$ distance in image space;
    \item[$\bullet$] $L_2$ on the deep features of the bottleneck layer of a custom denoising autoencoder;
    \item[$\bullet$] $L_2$ distance on the embeddings learned by traditional manifold learning LLE and Isomap;
    \item[$\bullet$] cosine distance of deep features of VGG and AlexNet. 
\end{itemize}

\subsubsection{Distance in Image Space}
To compute $L_2$ distance in image space, we represent each RGB image as a flat vector $\hat{x} \in \mathbb{R}^{w\times h \times c}$, where the our test images have $c=3$ color channels and a width $w=320$ and height $h=180$ pixels:

\begin{equation}
    d(x_i, x_j) = \sqrt{\sum_{c,w,h}\big(\hat{x}_i^{c,w,h}-\hat{x}_j^{c,w,h}\big)^2},
\end{equation}

\subsubsection{Distance in DAE Bottleneck Activation Space}
We compare our results with a custom denoising autoencoder (DAE) \cite{vincent2010stacked}. An autoencoder is a kind of neural network divided into two parts, an encoder and a decoder. We consider the output of the bottleneck layer as the features that the encoder retrieves and encodes from the input. The encoding network of our DAE reduces the dimension of each image $x$ to a lower dimensional latent vector $\hat{y} \in \mathbb{R}^{w\times h \times c}$, where the latent space has $c=500$ channel dimensions and a width $w=16$ and height $h=9$ spatial dimensions. To measure image similarity we use $L_2$ distance on the activations of the bottleneck layer as below:
\begin{equation}
    d(x_i, x_j) = \sqrt{\sum_{c,w,h}\big(\hat{y}_i^{c,w,h}-\hat{y}_j^{c,w,h}\big)^2}
\end{equation}
The architecture and training procedure of the denoising autoencoder is described in the supplementary materials.

\subsubsection{Distance in LLE and Isomap Embeddings}
The traditional manifold learning techniques LLE and Isomap map a set of images $X = \{x_i\}^m_{i=1}$ to a set of low dimensional vectors $Y = \{y_i\}^m_{i=1}$ where $y \in \mathbb{R}^d$ and $d\in \{1, \dots, m-1\}$ is the dimension of the embedding which must be specified by the user. In addition to the dimension of the embedding, the neighbors of each image must be specified.\par

In our comparison with traditional manifold learning, we test both LLE and Isomap with all parameters for the number of nearest neighbors, $k \in \{2, \dots, 10\}$ and embedding dimensions, $d \in \{2, \dots, 20\}$, with $L_2$ distance on the learned embedding vectors $Y^{k, d} = \{y^{k,d}_i\}^m_{i=1}$.
\begin{equation}
    d(x_i, x_j; k,d) = \sqrt{\sum_{c,w,h}\bigg(y^{k,d}_i - y^{k,d}_j\bigg)^2},
\end{equation}

\subsubsection{Cosine Distance in VGG and AlexNet Activation Space}
\begin{figure}[]
\centering
\subfloat[]{
  \includegraphics[width=55mm]{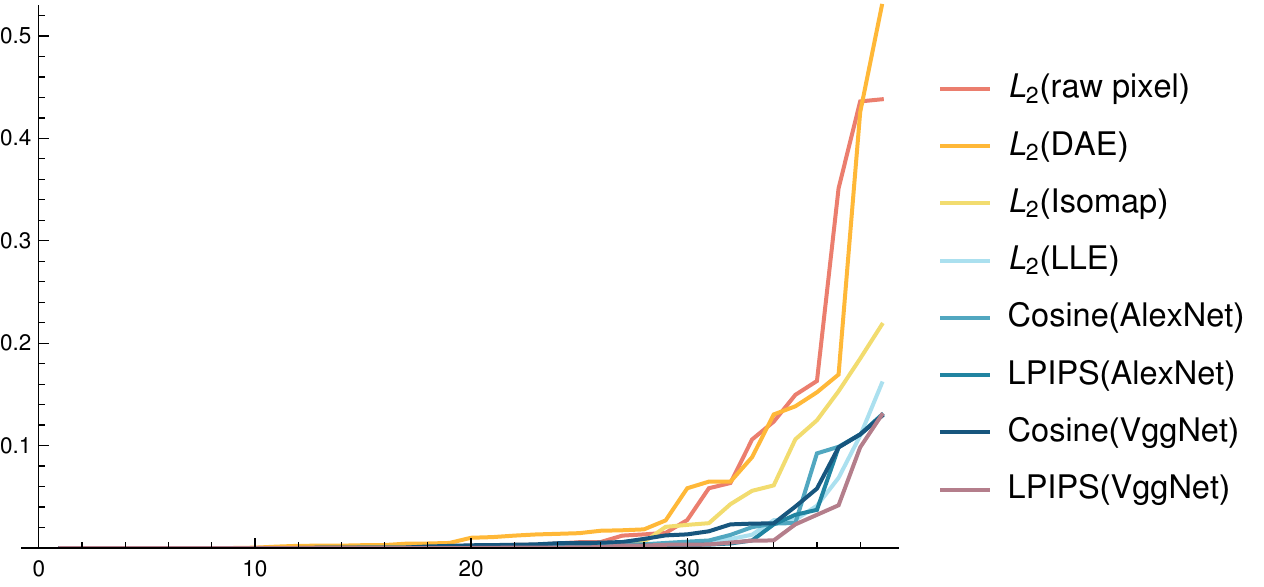}\hspace*{-0.3em}
}
\subfloat[]{
  \includegraphics[width=55mm]{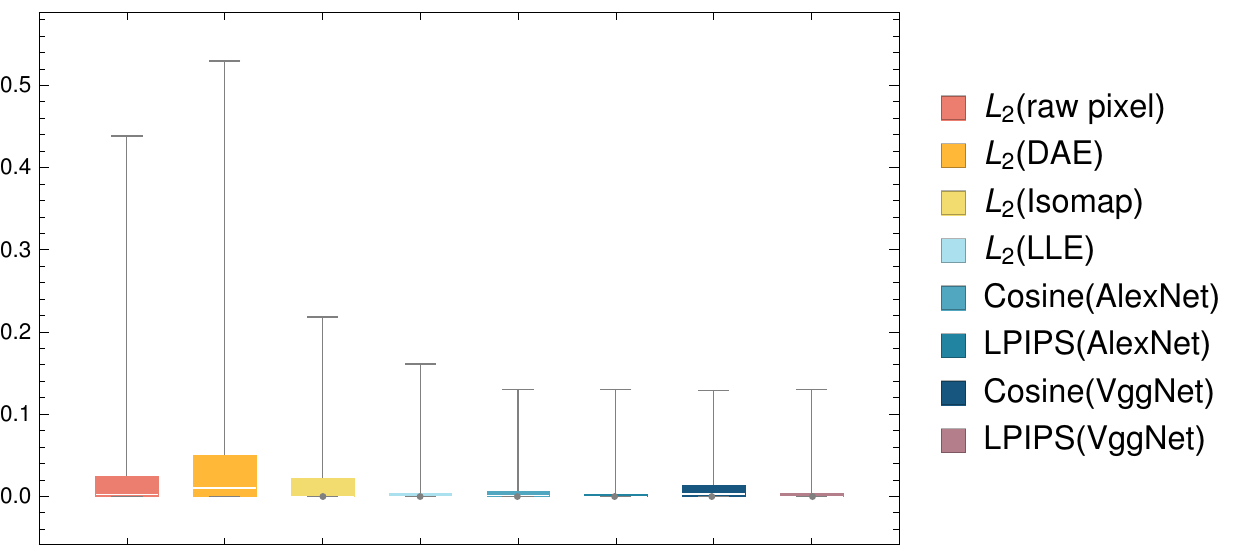}
}
\caption{\small (a) A comparison of the normalized Kendall tau distance for the reconstruction of 39 animations with different images similarity metrics (independently sorted). (b) A box and whisker chart for all test animations and test methods. The whisker endpoints show the maximum and minimum distance values, the solid box shows the 25 percent and 75 percent quantiles, and the white notch shows the median value.}
\label{fig_Kendall}
\end{figure}

Lastly we compare the cosine distance in the channel dimension of the same deep features used with the LPIPS metric described in Section 4.1.
\begin{equation}
    d(x_i, x_j) = \sum_l (1-\frac{1}{H_l W_l}\sum_{h,w} \hat{y}^l_{ihw}.\hat{y}^l_{jhw}),
\end{equation}
The Kendall tau distances for each method and each test case (sorted independently for clarity) and a box and whisker chart are shown in Figure \ref{fig_Kendall}.

\subsubsection{Animation Reconstruction Experiment}
For each image metric, we repeat the following procedure for each animation in our test set:
\begin{enumerate}
    \item compute the complete weighted graph of images with the appropriate distance function as edge weights;
    \item compute a Hamiltonian path from the node corresponding to the first frame to the node corresponding to the last frame;
    \item calculate the normalized Kendall tau distance \cite{kendall1938new} of the original sequence and the sequence generated by the Hamiltonian path.
\end{enumerate}

For an animation with $m$ frames, let $A = \langle X_i\rangle^m_{i=1}$ denote the original sequence of frames $X_i$ and let $H(A) = \langle X_{\psi(i)}\rangle^m_{i=1}$ denote the shortest Hamiltonian path from frame-$1$ to frame-$m$. Then the normalized Kendall tau distance between the original sequence and $H(A)$ is defined as:

\begin{equation}
    \kappa_\tau \big(H(A))\big) = \frac{2}{m(m-1)} \sum^m_{i=1}\sum^m_{j=i+1} \bar{\kappa}_\tau\big(\psi(i), \psi(j)\big),
\end{equation}
and
\begin{equation}
    \bar{\kappa}_\tau(i, j) = \begin{cases}
			0, & \text{if } i<j\\
            1, & \text{if } i>j,
		 \end{cases}
\end{equation}
The Kendall tau distance measures the number of discordant pairs in the Hamiltonian path sequence. It is normalized so that the distance $\bar{\kappa}_\tau(H(A)) \in [0,1]$ for any number of frames in the animation clip. A Hamiltonian path sequence with the same order as the test animation has zero distance and a sequence with the reverse order has a distance of one, thus the Normalized Kendall tau distance also gives a measure of rank correlation.\par

To consider an input animation as ground truth for a Hamiltonian path, it must not contain cyclic motion. Thus we visually inspect each animation and remove examples with cyclic motion. Additionally, we removed trivial cases where all test methods perfectly reconstruct the animation. \par

\begin{figure}[]
  \centering
  \includegraphics[width=0.7\textwidth]{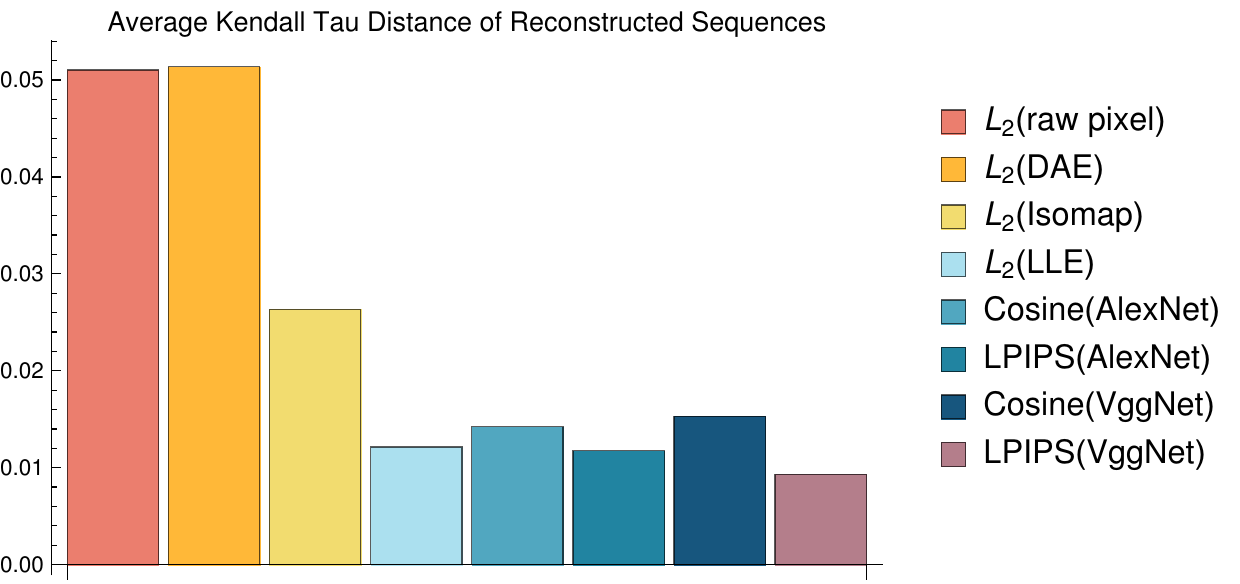}
  \caption{\small Comparison of the average Kendall tau distance (equation 12) for 39 reconstructed animations. We test 8 different distance measures, pairwise $L_2$ distance of the raw image pixels, the bottleneck layer of a denoising autoencoder, and the low dimensional embeddings learned by Isomap and LLE; cosine distance and LPIPS of the activations of selected layers of VGG and AlexNet.}
  \label{fig_6}
\end{figure}

In total, we tested the reconstruction of 39 animations. The average reconstruction errors are shown in Figure \ref{fig_6}. In the case of traditional manifold learning algorithms LLE and Isomap, we test all parameters $k \in \{2, \dots, 10\}$ and $d \in \{2, \dots, 20\}$ and select the lowest reconstruction error for each test animation. The results show that, on average, using features from VGG or AlexNet with the LPIPS metric produce Hamiltonian Path sequences which are closer to the original sequence than all other test metrics. While all similarity metrics have relatively small reconstruction errors, the bottleneck activations of the denoising autoencoder has the worst results. This may be due to the fact that the DAE is trained solely on japanese manga style images. More diverse training data could possibly improve the results of the DAE’s bottleneck features. The traditional manifold learning technique LLE outperforms Isomap and slightly outperforms the cosine distance of the extracted deep features of VGG and Alexnet, however the experiment was slightly biased towards traditional manifold learning since each animation was tested with 171 different parameter settings and only the single best result was counted towards the average reconstruction error. Despite this bias, LPIPS with VGG features and AlexNet features performed the better than LLE without the need for any parameter tuning. We give additional results and details of the reconstruction experiment in our supplementary material.

\subsection{Additional Applications}
\subsubsection{Image Layouts}
The proposed framework can also be used to create image layouts used for quickly browsing large collections of unordered images. Placing perceptually similar images next to each other can improve human image retrieval tasks by reducing the perceptual load and thus accelerating visual processing \cite{schoeffmann2011similarity}.\par

\begin{figure}[]
  \centering
  \includegraphics[width=0.8\textwidth]{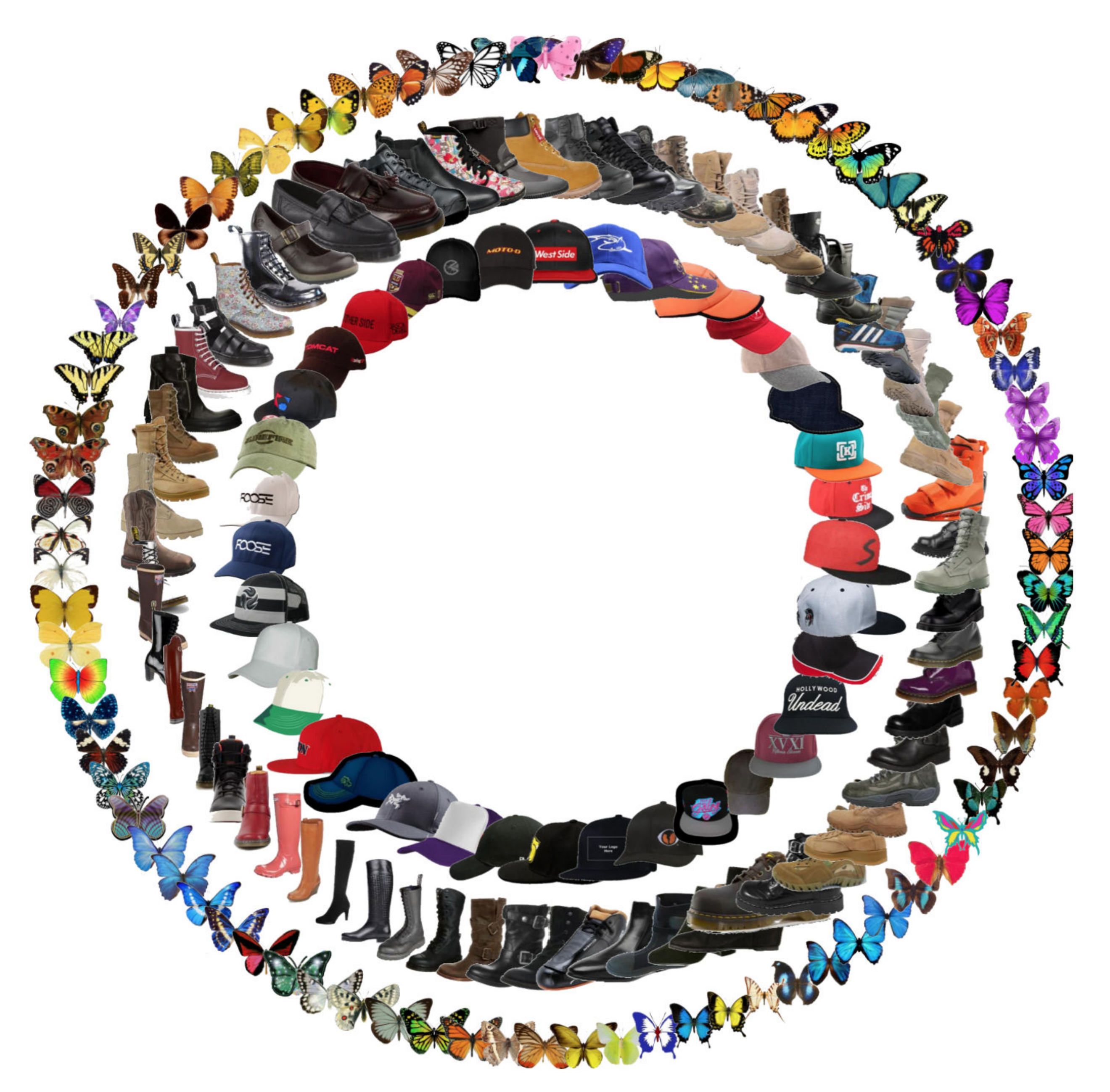}
  \caption{\small A radial image layout with sequences automatically generated by our system.}
  \label{fig_radial_layout}
\end{figure}

\begin{figure}[]
  \centering
  \includegraphics[width=0.95\textwidth]{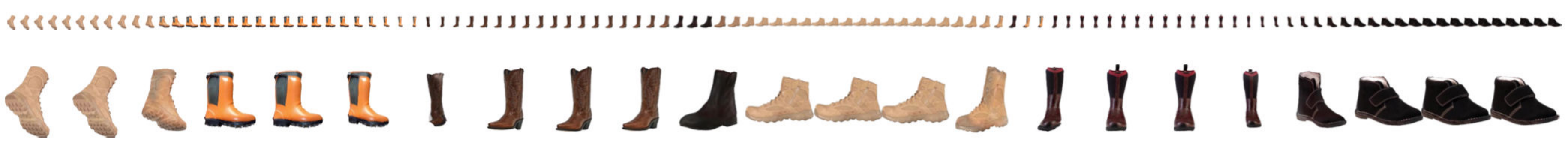}
  \caption{\small Linear image layout example generated by the proposed method. Readers are suggested to see our supplementary video for a better visualization.}
  \label{fig_linear_layout}
\end{figure}

We tested our framework on input data for the data driven morphing technique proposed by \cite{averbuch2016smooth} comprised of four image sets where each image set contains between 148 and 722 images of different instances of the same object. Our framework was capable of producing many smooth and visually appealing image layouts from this data. Because of a large number of images in the datasets, our framework can be useful for visualizing smooth sub-sequences. For example, by identifying continuous subsequences of a given length with a minimum perceptual distance between adjacent frames or sampling longer sequences at even intervals. Figure \ref{fig_radial_layout} shows a radial image layout for images sequences generated with our proposed method, an example of how our system could be used to visualize large datasets of images of similar objects. Figure \ref{fig_linear_layout} shows an example of a smooth linear image layout generated by the proposed Hamiltonian path sequencing method applied to a collection of textured boot images. In the supplementary materials, we also present more results on this kind of application.

\begin{figure}[]
  \centering
  \includegraphics[width=0.85\textwidth]{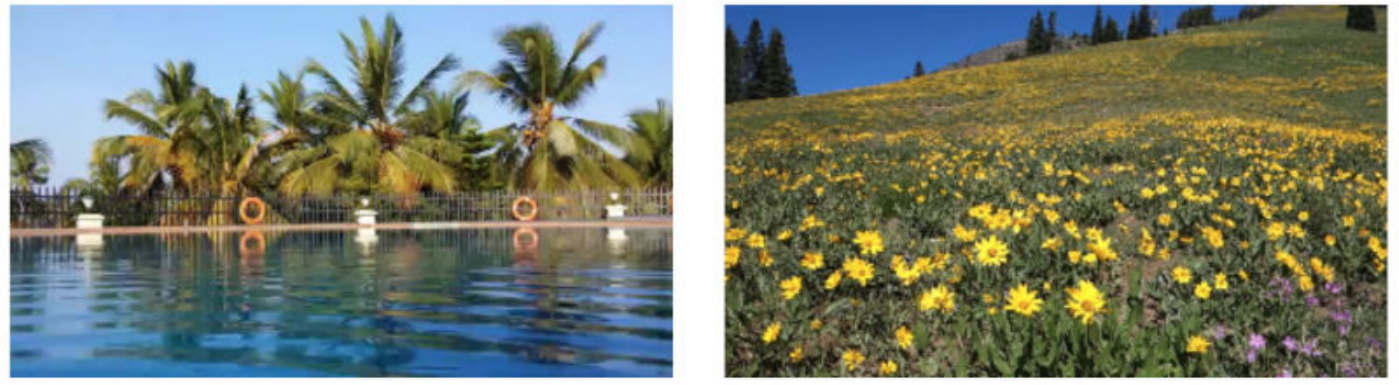}
  \caption{\small Natural image video examples.}
  \label{fig_synthesis}
\end{figure}

\subsubsection{Video Synthesis}
While this work focuses on animation video resequencing, our framework also applies to natural image video resequencing. If the input images depict stochastic motion, such as grass swaying in the win or ripples of water in a pool as the examples shown in Figure \ref{fig_synthesis}, smooth video resequences and cycle animations can be generated using the LPIPS distance and the graph traversal algorithm described in our framework.

\subsection{Limitations}
\begin{figure}[]
\centering
\subfloat{
  \includegraphics[width=55mm]{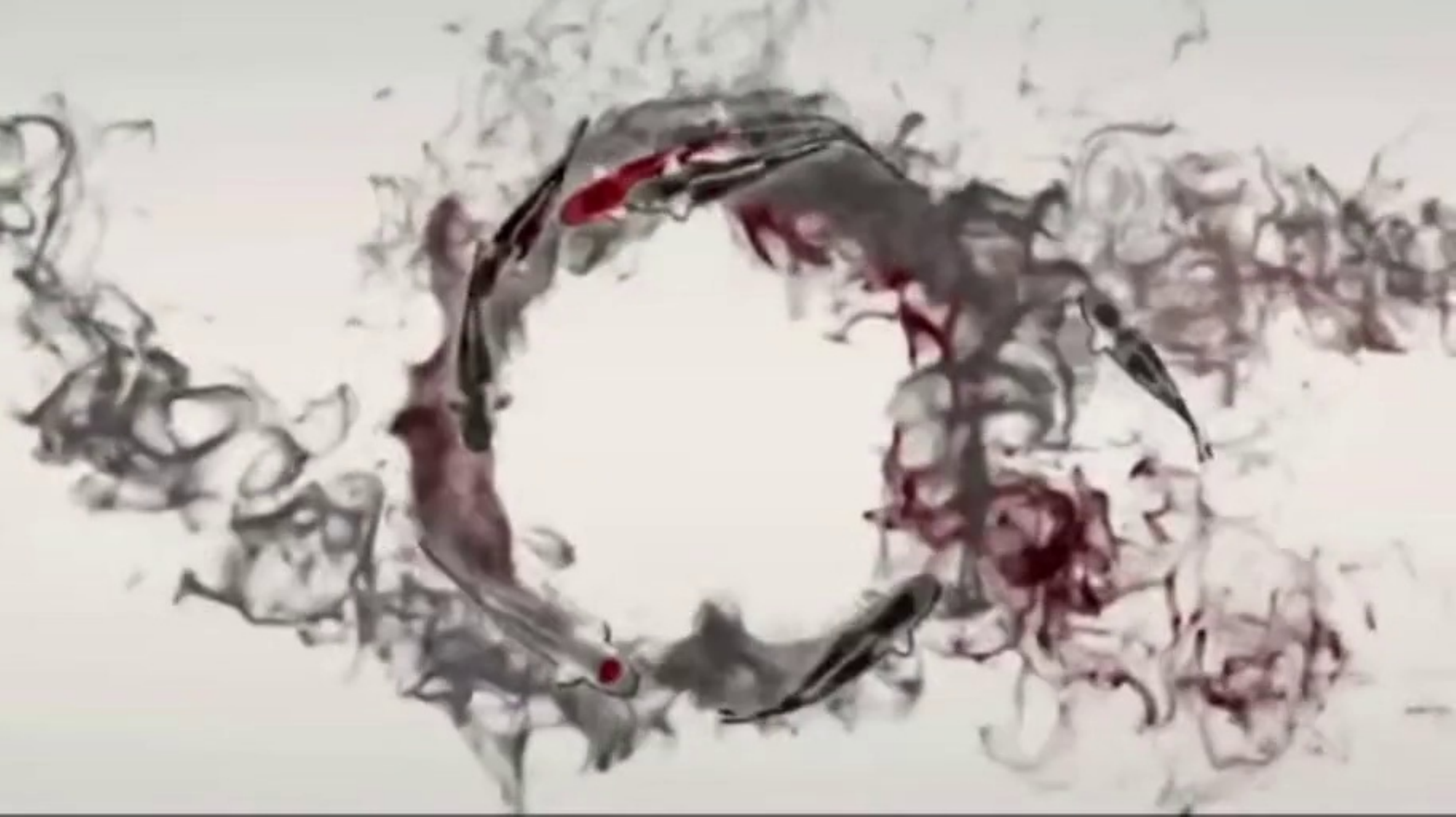}\hspace*{-0.3em}
}
\subfloat{
  \includegraphics[width=55mm]{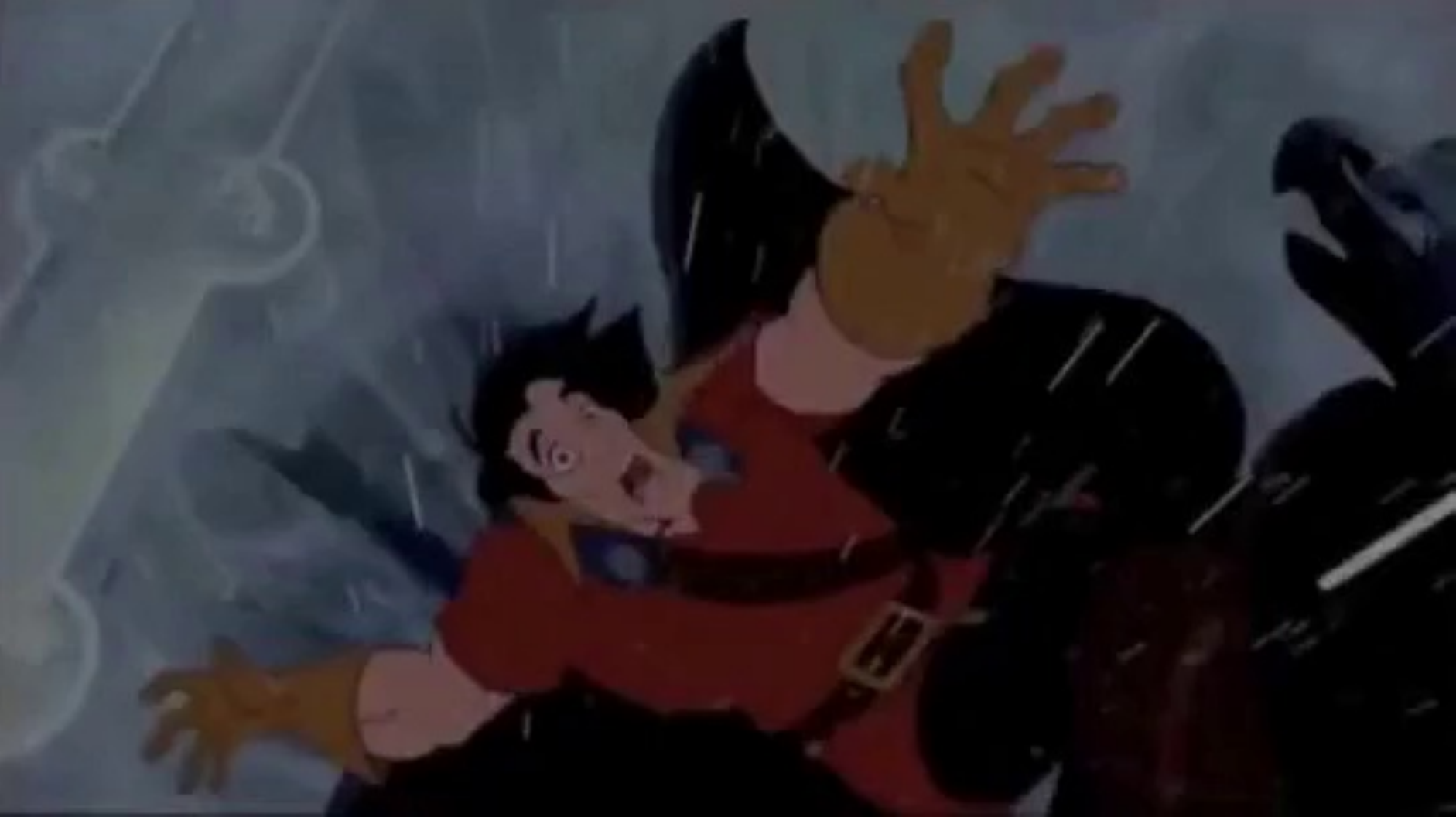}
}
\caption{{\small The school of fish example (a) and the falling sequence (b) are the examples where the proposed framework cannot produce smooth sequences other than the original video sequence.}}
\label{fig_limitation}
\end{figure}

Our framework’s main limitation is its dependence on the input data. If the collection of input images is taken by densely sampling a video sequence with a strong distinction between backward and forward motion, such as the school of fish and falling sequence shown in Figure \ref{fig_limitation}, then the MST may be path-like, and the proposed sequencing methods will likely select frames which are very similar to the dynamics of the original video sequence. In general, it may not be possible to generate new dynamics from input collections that do not contain a sufficient variety in motion and appearance. One possible way to overcome this limitation would be to develop an image-synthesis technique to generate new images that interpolate or extrapolate new motion by considering motion directions of objects.

\section{Conclusion and Future work}
We proposed a novel deep-learning framework for a new application for animation video resequencing which can generate smooth sequences and subsequences for many image styles. Our framework can serve as an efficient tool to automatically create new animation sequences from a collection of images. We also believe our framework could assist users in creating a comprehensive animation dataset by extracting many smooth subsequences from existing animation data. To our knowledge, a well-labeled dataset for general animation data does not yet exist.\par

Our results suggest that the activations of convolutional neural networks are useful features for smooth sequencing of photorealistic, non-photorealistic, ordered, and unordered image collections. Our quantitative analysis shows that deep-features and the LPIPS metric can reconstruct animation sequences with greater accuracy than cosine distance of the same features, $L_2$ distance of the activations of a denoising autoencoder, $L_2$ distance in image space, and $L_2$ distance in the embedding space obtained by traditional manifold learning. Our qualitative results also show that the LPIPS metric produces a visible improvement over these other methods.\par

Despite the various styles, animators utilize a standard set of principles, including natural movement, to create more realistic looking animations. Thus, in the future, we would like to develop a self-supervised learning technique to extract motion features from existing animation video and combine metric learning and sequencing in a single deep learning optimization framework to solve problems in Figure \ref{fig_limitation}.

\bibliographystyle{abbrv}
\bibliography{references}
%\nocite{*} % to test all bib entrys
%\bibliographystyle{unsrt}
%\bibliography{references} % file mwe.bib
\end{document}